\documentclass[english,reprint, longbibliography, superscriptaddress, breaklinks=true, showkeys, showpacs=false, nofootinbib]{revtex4}
\usepackage[T1]{fontenc}
\usepackage[latin9]{inputenc}
\usepackage{color}
\usepackage{babel}
\usepackage{amsmath}
\usepackage{amssymb}
\usepackage{graphicx}
\usepackage{physics}
\usepackage{subfigure}
\usepackage{caption}
\usepackage[unicode=true,pdfusetitle,
 bookmarks=true,bookmarksnumbered=false,bookmarksopen=false,
 breaklinks=true,pdfborder={0 0 0},backref=false,colorlinks=true]
 {hyperref} 
\usepackage{breakurl}

\makeatletter
\@ifundefined{textcolor}{}
{%
 \definecolor{BLACK}{gray}{0}
 \definecolor{WHITE}{gray}{1}
 \definecolor{RED}{rgb}{1,0,0}
 \definecolor{GREEN}{rgb}{0,1,0}
 \definecolor{BLUE}{rgb}{0,0,1}
 \definecolor{CYAN}{cmyk}{1,0,0,0}
 \definecolor{MAGENTA}{cmyk}{0,1,0,0}
 \definecolor{YELLOW}{cmyk}{0,0,1,0}
}

\pdfoutput=1
\hypersetup{colorlinks=true,citecolor=blue,linkcolor=cyan,urlcolor=blue,filecolor= green, breaklinks=true}
\usepackage{url}
\usepackage{breakurl}

\makeatother

\begin{document}

\title{Complete complementarity relations for multipartite pure states}

\author{Marcos L. W. Basso}
\email{marcoslwbasso@mail.ufsm.br}
\address{Departamento de F\'isica, Centro de Ci\^encias Naturais e Exatas, Universidade Federal de Santa Maria, Avenida Roraima 1000, Santa Maria, RS, 97105-900, Brazil}

\author{Jonas Maziero}
\email{jonas.maziero@ufsm.br}
\address{Departamento de F\'isica, Centro de Ci\^encias Naturais e Exatas, Universidade Federal de Santa Maria, Avenida Roraima 1000, Santa Maria, RS, 97105-900, Brazil}

\selectlanguage{english}%

\begin{abstract}
Complementarity relations for wave-particle duality are saturated only for pure, single-quanton, quantum states. For a completely incoherent state, it is known that wave and particle quantifiers can reach zero, and hence no information about the wave and particle aspects of the system can be obtained. This means that the information is being shared with another systems, and quantum correlations can be seen as responsible for the loss of purity of the quanton, provided that the quanton is part of a multipartite pure quantum system. In this paper, by exploring the purity of bi- and tri-partite pure quantum states, we show that it is possible to obtain complete complementarity relations. This procedure allows us to create a general framework for obtaining complete complementarity relations for a subsystem that belongs to an arbitrary multi-partite quantum system in a pure state. Besides, by some simple examples, we show that if the predictability measure is changed then the correlation measure must also be changed in order to obtain complete complementarity relations for pure cases.
\end{abstract}

\keywords{Wave-particle duality; Multipartite quantum systems; Complete complementarity relations}

\maketitle

\section{Introduction}

\label{intro}
One of the most intriguing aspects of quantum mechanics is the wave-particle duality. This aspect is generally captured, in a qualitative way, by Bohr's  complementarity principle \cite{Bohr}. It states that quantons \cite{Leblond} have characteristics that are equally real, but mutually exclusive. The wave-particle duality is the best known example of this principle, where, in a two-way interferometer, such as the Mach-Zehnder interferometer or the double-slit interferometer,  the wave aspect is characterized by the interference fringes visibility, while the particle nature is given by the which-way information of the path along the interferometer, such that the complete knowledge of the path destroys the interference pattern visibility and vice-versa. The first quantitative version of the wave-particle duality was explored by Wootters and Zurek \cite{Wootters}, when they investigated interferometers in which one obtains incomplete which-way information by introducing a path-detecting device, and showed that a sharply interference pattern can still be retained. Later, this work was extended by Englert, who derived a wave-particle duality relation \cite{Engle}. Also, using a different line of reasoning, Greenberger and Yasin \cite{Yasin}, considering a two-beam interferometer wherein the intensity of each beam was not necessarily the same, defined a measure of path information called predictability. Hence, if the quantum system passing through the beam-splitter has different probabilities of getting reflected in the two paths, one could predict the path information of the quantum system better than when those probabilities are equal. This kind of reasoning was followed by Jaeger, Shimony, and Vaidman \cite{Jaeger}, and can be summarized by a simple complementarity relation:
\begin{equation}
    P^2 + V^2 \le 1 \label{eq:cr1},
\end{equation}
where $P$ is the path predictability and $V$ is the visibility of the interference pattern. It is worth noticing that these aspects of a quantum system are not necessarily mutually exclusive. An experiment can provide partial information about the wave and particle nature of a quantum system, but the more information it gives about one aspect of the system, the less information the experiment can provide about the other. More recently, several steps have been taken towards the quantification of the wave-particle duality by many authors, such as D\"urr \cite{Durr}  and Englert et al. \cite{Englert}, that established criteria for checking the reliability of newly defined predictability measures and interference pattern quantifiers, and extended measures of the wave-particle aspects to discrete $d$-dimensional quantum systems. As well, with the development of the field of quantum information, it was suggested that the quantum coherence \cite{Baumgratz} would be a good generalization of the visibility measure \cite{Bera, Bagan, Tabish, Mishra}. Meanwhile, the predictability is a measure of the knowledge about the quantum level wherein a quanton can be found. These levels can represent, besides the paths on a Mach-Zehnder interferometer, energy levels of an atom \cite{Xu} or, more generally, population levels \cite{Vlatko}. So far, many lines of thought were taken for quantifying the wave-particle properties of a quantum system \cite{Angelo, Coles, Hillery, Qureshi, Maziero}. 

However, complementarity relations like the one in Eq. (\ref{eq:cr1}) are saturated only for pure, single-particle, quantum states. For mixed states, the left hand side is always less than one and can even reach zero for a maximally mixed state. Hence no information about the wave and particle aspects of the system can be obtained. As noticed by Jakob and Bergou \cite{Janos}, this lack of knowledge about the system is due to another intriguing quantum feature: entanglement \cite{Bruss, Horodecki}. This means that the information is being shared with another system, and this kind of quantum correlation can be seen as responsible for the loss of purity of each subsystem such that, for pure maximally entangled states, it is not possible to obtain information about the local properties of the subsystems. As showed by these authors, the concurrence \cite{Woot} is recognized as the appropriate quantum correlation measure in a bipartite state of two qubits that completes relation (\ref{eq:cr1}). It is noteworthy that this complete relation for two qubits was claimed to be experimentally confirmed recently \cite{Qian}. Jakob and Bergou extended this idea for composite bipartite systems of arbitrary dimension \cite{Jakob, Bergou}, suggesting that there must exist a complementary relation between the information of the local properties of each subsystem and the entanglement of the composite system, and showed that I-Concurrence \cite{Rungta} is the measure of quantum correlation that completes relation (\ref{eq:cr1}) for composite bipartite pure states. Following the same reasoning, Hiesmayr and Huber \cite{Huber} derived an operational entanglement measure for any multiparty system.    

However, it's known that entanglement is not the only quantum correlation existing in multipartite quantum systems \cite{Bromley, Adesso, Xi, Xiz, Yao}. For example, quantum discord is a type of quantum correlation that describes the incapacity of obtaining information from one interacting subsystem without perturbing it \cite{Zurek}. Quantum coherence in a composite system can be contained either locally or in the correlations between the subsystems. The portion of quantum coherence contained within correlations can be viewed as a kind quantum correlation, called correlated coherence \cite{Tan, Kraft}. Therefore a natural question one could ask, in the context of complementarity relations, is whether entanglement measures are the only quantum correlation measures that complete relations like (\ref{eq:cr1}). In this article we show that the answer to this question is negative. By exploring some simple examples, we show that if one changes the predictability measure, one has to change also the correlation measure in order to obtain a complete complementarity relation for pure multipartite states. Also, we show that, by exploring the purity of bipartite quantum systems, one can obtain a complete complementarity relation equivalent to the one obtained in \cite{Jakob, Bergou}, and a new one if we use the relative entropy of quantum coherence as measure of visibility. In addition, by exploring the purity of tripartite quantum systems, one can obtain a complete complementarity relation with a measure of correlation equivalent to the generalized concurrence of tripartite systems, obtained in \cite{Bhaskara} using exterior algebra. This procedure allowed us to extend the work made of Jakob and Bergou as we give a general framework where the complete complementarity relations appear naturally from the purity of a multipartite quantum system. In addition, it's worth pointing out that in \cite{Huber} the authors explored the purity-mixedness relation of a quanton to obtain a multidimensional complementarity relation, which was used to define an operational entanglement measure. In this article, we obtain the same complete complementarity relation  (and associated measures of predictability, visibility, and quantum correlations) within the purity of the multipartite quantum system. In this way, we were able to express the complementarity measures in terms of the elements of the density matrix of the whole system, once the state that is known (in principle) is the state of the multipartite quantum system, and not only the state of a particular subsystem. Besides, Tessier \cite{Tessier} studied the relation between the local properties of a quanton and entanglement, and conjectured that such relation is a general feature of composite multipartite quantum systems of arbitrary dimension. Hence, our work proves the conjecture made by Tessier.

We organized the remainder of this article in the following manner. In Sec. \ref{sec:bi}, we explore the properties of bipartite pure quantum systems, and, using the relative entropy of coherence, we obtain complementarity relations equivalent to those reported in \cite{Jakob, Bergou} and new ones. Also, exploring the purity of a tripartite quantum system, we obtain a new complete complementarity relation in Sec. \ref{sec:tri}. Next, we obtain complete complementarity relations from the purity of an arbitrary multipartite quantum system in Sec. \ref{sec:multi}. In Sec. \ref{sec:rqc}, we show that in changing the predictability measure, one also has to change the correlation measure in order to obtain a complete complementarity relation for pure states. Finally, we give our conclusions in Sec. \ref{sec:conc}.

\section{Complementarity Relations for Bipartite Pure Quantum Systems }
\label{sec:bi}
\begin{figure}[htp]
\subfigure[\normalsize $C_{hs}$ as function of $x$ for some values of $w$.]{\includegraphics[scale = 0.38]{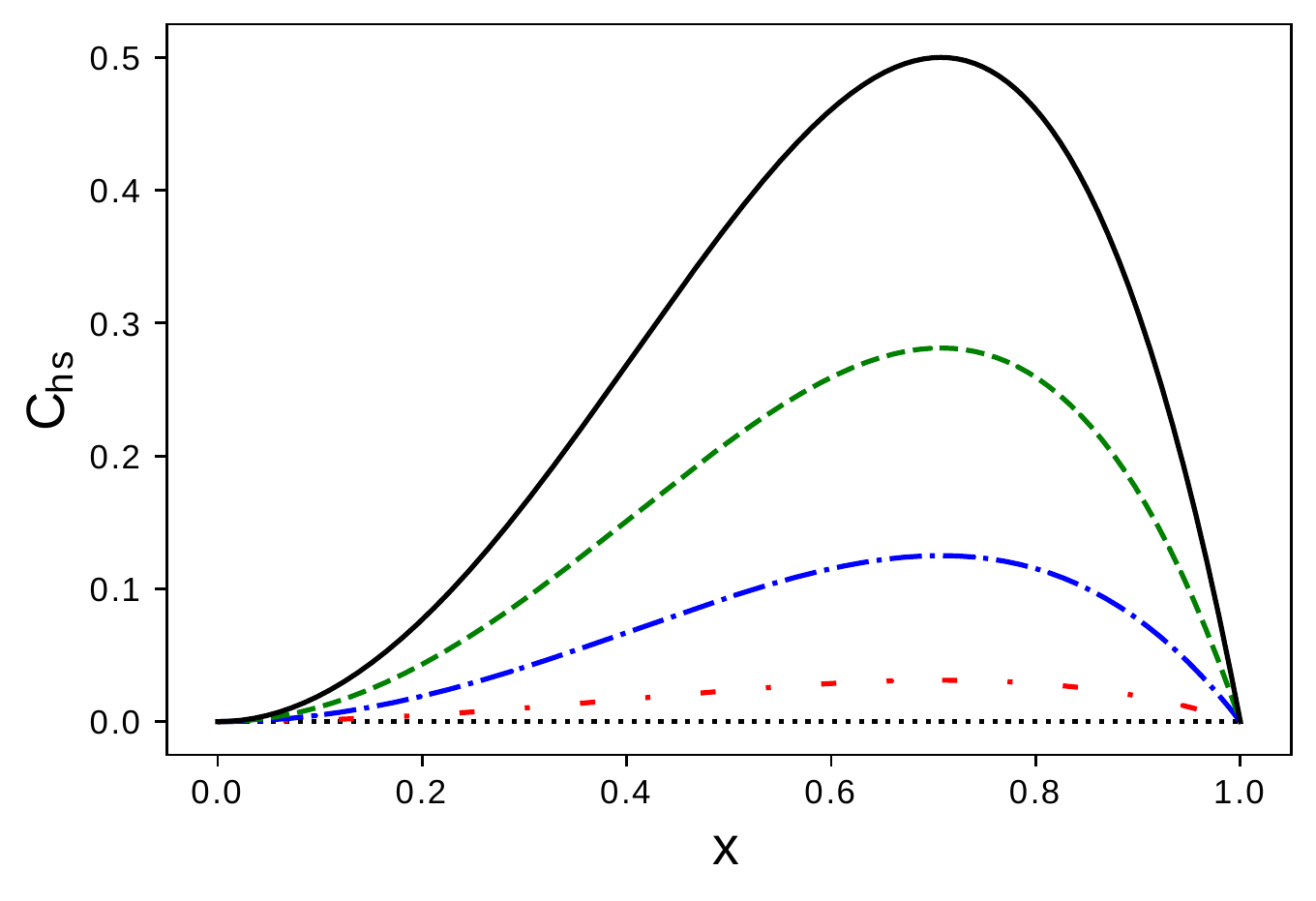}{\label{fig:a}}}
\subfigure[\normalsize $P_{hs}$ as function of $x$ for some values of $w$.]{\includegraphics[scale = 0.38]{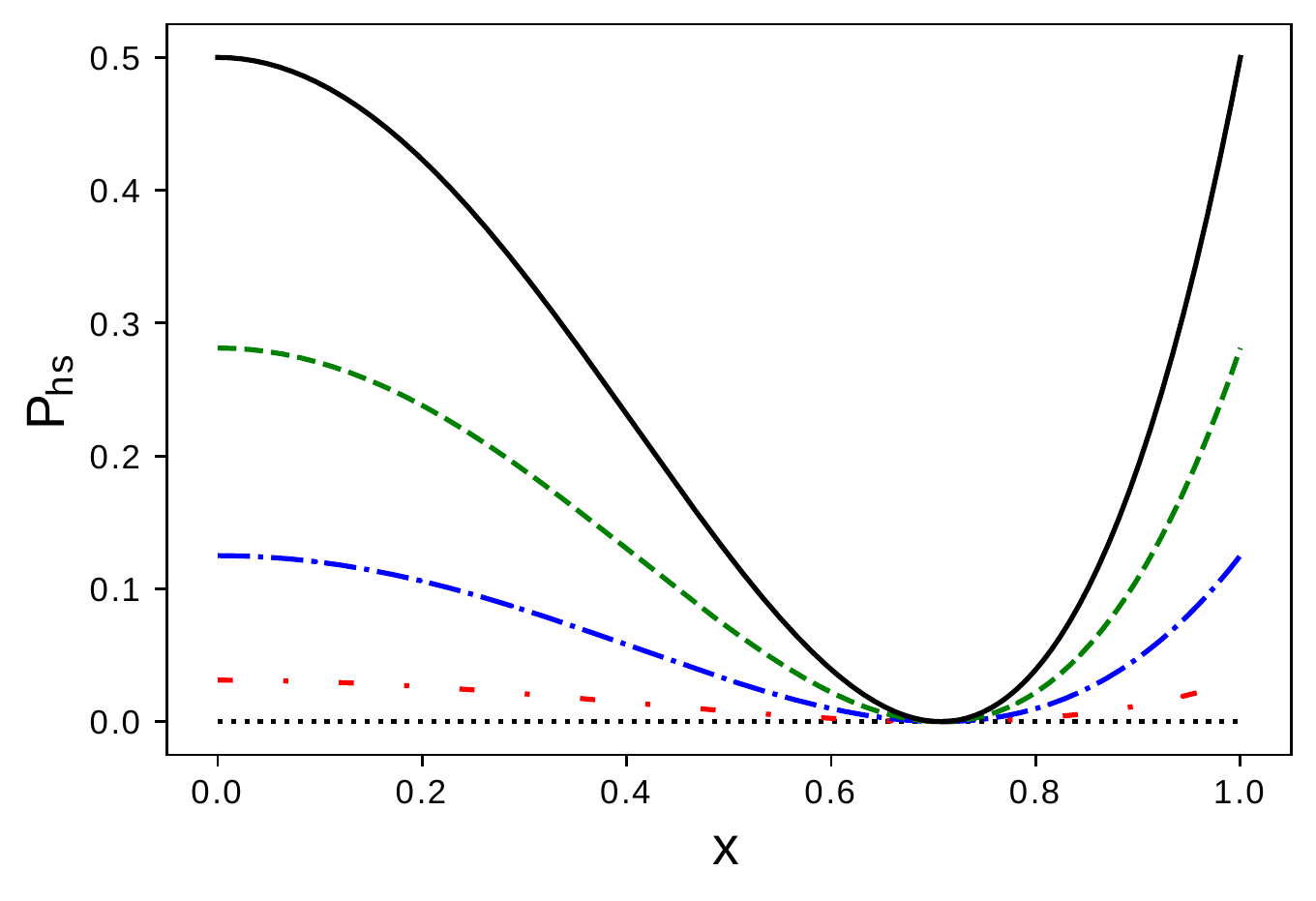}{\label{fig:b}}}
\qquad
\subfigure[\normalsize $P_{hs} + C_{hs}$ as function of $x$ for some values of $w$.]{\includegraphics[scale = 0.38]{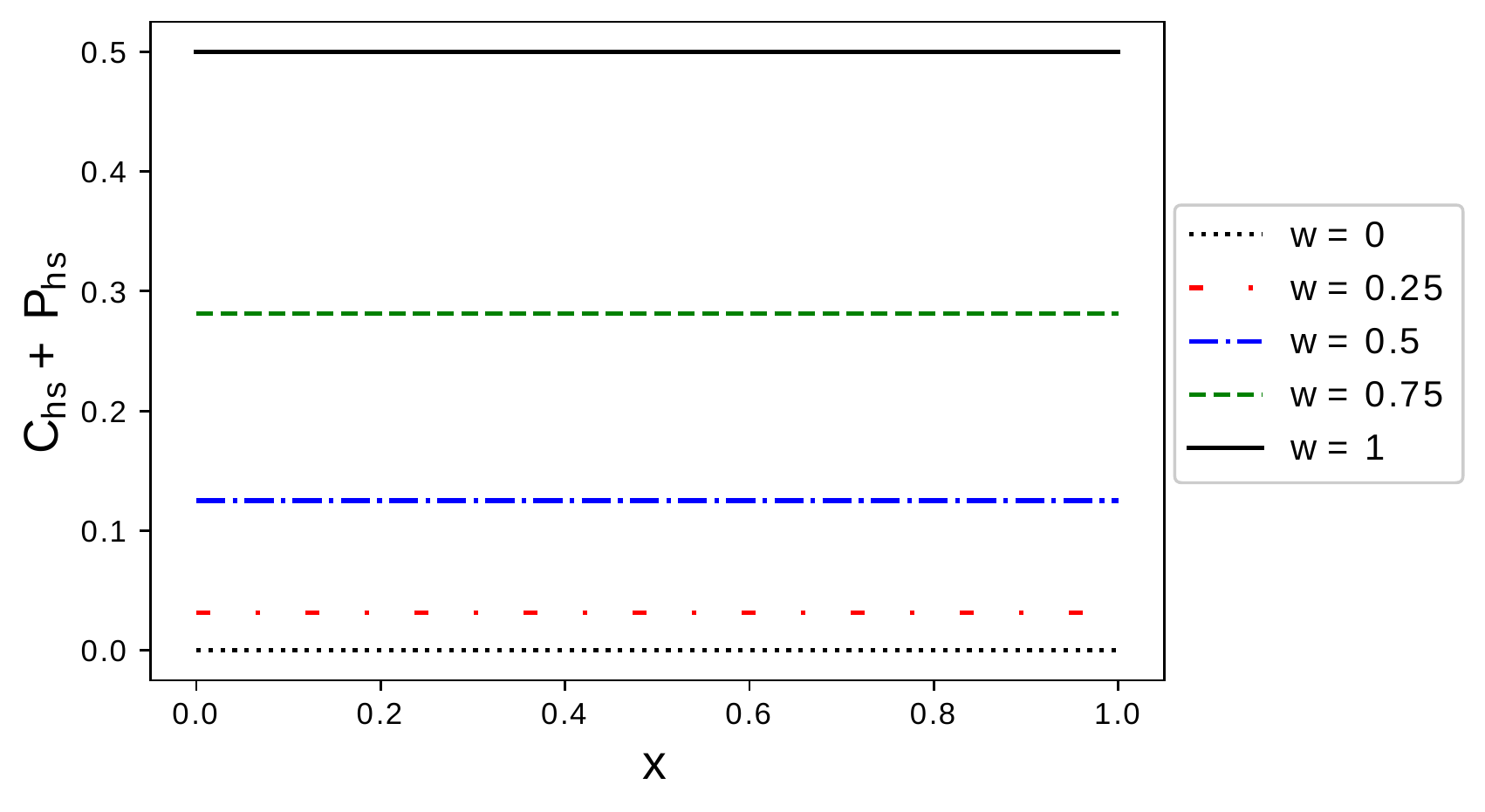}{\label{fig:c}}}
\caption{\normalsize Balanced exchange between predictability and coherence of the state in Eq. (\ref{eq:werner}). }
\label{fig:wstate}
\end{figure}

In this section, we will explore the properties of bipartite pure quantum systems and their respective subsystems $A$ and $B$ of dimension $d_A$ and $d_B$, respectively. So, a general state of the system $ \ket{\Psi}_{A,B}$ can be represented as a vector in the composite Hilbert space $\mathcal{H}_{A} \otimes \mathcal{H}_{B}$ with dimension $d = d_A d_B$ \cite{Mark}. Let $\{\ket{i}_A\}_{i = 0}^{d_A - 1}$, $\{\ket{j}_B\}_{j = 0}^{d_B - 1}$ be a local orthonormal basis for the spaces $\mathcal{H}_{A}$, $\mathcal{H}_{B}$, respectively, so that $\{\ket{i}_A \otimes \ket{j}_B := \ket{i,j}_{A,B}\}_{i,j = 0}^{d_A - 1, d_B - 1}$ is an orthonormal basis for representing vectors in $\mathcal{H}_{A} \otimes \mathcal{H}_{B}$. Therefore, an arbitrary state of a bipartite quantum system can be written as $  \ket{\Psi}_{A,B}= \sum_{i,j = 0}^{d_A - 1, d_B - 1} a_{ij} \ket{i,j}_{A,B}$, or, equivalently, by the density operator \cite{Messiah, Fano}
\begin{equation}
    \rho_{A,B} = \sum_{i,k=0}^{d_A - 1}\sum_{j,l=0}^{d_B - 1 }\rho_{ij,kl} \ket{i,j}_{A,B}\bra{k,l},
\end{equation}
where $\rho_{ij,kl} = a_{ij}a^*_{kl}$. Meanwhile, the state of subsystem A (B) are obtained by tracing over B (A):
\begin{align}
    & \rho_A = \sum_{i,k=0}^{d_A - 1} \rho^A_{ik} \ket{i}_A\bra{k} = \sum_{i,k=0}^{d_A - 1}\sum_{j = 0}^{d_B - 1}\rho_{ij,kj} \ket{i}_A\bra{k},\\ 
    & \rho_B = \sum_{j,l=0}^{d_B - 1} \rho^B_{jl} \ket{j}_B\bra{l} =  \sum_{j,l=0}^{d_B - 1}\sum_{i = 0}^{d_A - 1}\rho_{ij,il} \ket{j}_B\bra{l}.
\end{align}
In general, the states of the subsystem A and B are not pure, which implies that some information of the subsystems is missing. It's easy to see this fact by exploring the properties of the density matrix of one of the subsytems. For example, once the state of the subsystem A is mixed, we have $1 - \Tr \rho_A^2 >0$, or equivalently,
\begin{align}
    1 - \sum_{i,k = 0}^{d_A - 1}\abs{\sum_{j = 0}^{d_B - 1}\rho_{ij,kj}}^2 > 0 \label{eq:ine},
\end{align}
which can be written as a complementarity relation obtained in \cite{Maziero}:
\begin{align}
    P_{hs}(\rho_A) + C_{hs}(\rho_A) < \frac{d_A - 1}{d_A}, \label{eq:phs} 
\end{align}
where $P_{hs}(\rho_A) = \sum_{i=0}^{d_A - 1} (\rho^A_{ii})^2 - 1/d_A = \sum_{i=0}^{d_A - 1}(\sum_{j = 0}^{d_B - 1}\rho_{ij,ij})^2  - 1/d_A$ is the predictability measure and $C_{hs}(\rho_A) =  \sum_{i \neq k = 0}^{d_A - 1}\abs{\rho^A_{ik}}^2 = \sum_{i \neq k = 0}^{d_A - 1} \abs{\sum_{j = 0}^{d_B - 1}\rho_{ij,kj}}^2 $ is the Hilbert-Schmidt\footnote{Even though $C_{hs}(\rho_A)$ is not considered a coherence monotone, it's a bona fide measure for the visibility of a quanton \cite{Maziero}.} quantum coherence \cite{Jonas}. The information content absent from system A is represented by the strict inequality in equation (\ref{eq:ine}) and (\ref{eq:phs}). For instance, if we consider the following state 
\begin{equation}
    \rho_A = w \ketbra{\psi} + \frac{1- w}{2} I_{2 \times 2}, \nonumber
    \label{eq:werner}
\end{equation}
with $\ket{\psi} = x \ket{0} + \sqrt{1 - x^2}\ket{1}$, $x \in [0,1]$, one can observe that for $w = 0$ we have a maximally incoherent state $\rho_A = \frac{1}{2}I_{2 \times 2}$, and no information about the quanton can be obtained, since $P_{hs} = C_{hs} = 0$. As pointed out by Qian et al. \cite{Qian}, the complementarity relation shown above do not really predict a balanced exchange between $C_{hs}$ and $P_{hs}$ simply because the inequality permits a decrease of $C_{hs}$ and $P_{hs}$ together, or an increase by both. It even allows the extreme case $P_{hs} = C_{hs} = 0$ to occur (neither wave or particle) while, in an experimental setup, we still have a quanton on hands. Thus, one can see that something must be missing from equations similar to Eq. (\ref{eq:phs}). In Fig. \ref{fig:wstate}, we plotted the balanced exchange between $C_{hs}$ and $P_{hs}$ for different values of $w$. By inspecting Fig. \ref{fig:wstate}, we can see that as $w \to 0$ the measures of predictability and visibility never reach their maximum possible values and $P_{hs} + C_{hs} \to 0$. As we'll see, what is missing for completely quantifying the behavior of a quanton is its correlation with other systems.

For convenience, from now on we'll sometimes omit the upper limits of the summations. As pointed out by Jakob and Bergou \cite{Jakob, Bergou}, the missing information is being shared via the correlations with the subsystem B, and we can see this by exploring the purity of the bipartite quantum system:
\begin{align}
   1 - \Tr \rho_{A,B}^2= 1 - \Big(\sum_{\overset{i = k}{j = l}} + \sum_{\overset{i \neq k}{j \neq l}} + \sum_{\overset{i \neq k}{j = l}} + \sum_{\overset{i = k}{j \neq l}}\Big)\abs{\rho_{ij,kj}}^2 = 0,
\end{align}
which can be rewritten as a complete complementarity relation for both subsystems
\begin{align}
    & P_{hs}(\rho_A) + C_{hs}(\rho_A) + C^{nl}_{hs}(\rho_{A|B}) = \frac{d_A - 1}{d_A} \label{eq:cr2},\\
    & P_{hs}(\rho_B) + C_{hs}(\rho_B) + C^{nl}_{hs}(\rho_{B|A}) = \frac{d_B - 1}{d_B} \label{eq:cr3},
\end{align}
given that we define the non-local quantum coherence of A (B) (that is being shared with B(A)) as
\begin{align}
    & C^{nl}_{hs}(\rho_{A|B}) := \sum_{\overset{i \neq k}{j \neq l}}\abs{\rho_{ij,kl}}^2 - 2 \sum_{\overset{i \neq k}{j < l}}\mathbb{R}e (\rho_{ij,kj} \rho^*_{il,kl}), \\
    & C^{nl}_{hs}(\rho_{B|A}) := \sum_{\overset{i \neq k}{j \neq l}}\abs{\rho_{ij,kl}}^2 - 2 \sum_{\overset{i < k}{j \neq l}}\mathbb{R}e (\rho_{ij,il} \rho^*_{kj,kl}).
\end{align}
The complementarity relations (\ref{eq:cr2}) and (\ref{eq:cr3}) are equivalent to the relations $P^2_k + V^2_k + [C^{(d)}_{AB}]^2 \le \frac{2(d_k - 1)}{d_k}$, $k = A,B$, for pure states, obtained by Jakob and Bergou in \cite{Jakob, Bergou}, where $C^{(d)}_{AB} = \sqrt{2(1 - \Tr \rho_k^2)}, \ k = A,B$, is the generalized concurrence. To see this, it is enough to notice that $1 - \Tr \rho_k^2 = C^{nl}_{hs}(\rho_{k|l})$ for $k \neq l = A,B$, $P_{hs}(\rho_k) = \frac{1}{2}P_k^2$, and $C_{hs}(\rho_k) = \frac{1}{2}V_k^2$ for $k = A, B$. Also, $C^{nl}_{hs}(\rho_{k|l}) \ge 0$ with $k \neq l = A,B$, for all bipartite pure quantum systems.

For any quantum state $\rho_A$ of dimension $d_A$, the relative entropy of coherence is defined as  \cite{Baumgratz}
\begin{align}
    C_{re}(\rho_A) = \min_{\iota \in I} S_{vn}(\rho_A||\iota),
\end{align}
where $I$ is the set of all incoherent states, and $S_{vn}(\rho_A||\iota) = \Tr(\rho_A \ln \rho_A - \rho_A \ln \iota)$ is the relative entropy. The minimization procedure leads to $\iota = \rho_{Adiag} = \sum_{i = 1}^{d_A} \rho^A_{ii} \ketbra{i}$. Thus 
\begin{align}
    C_{re}(\rho) = S_{vn}(\rho_{Adiag}) - S_{vn}(\rho_A) \label{eq:cre}.
\end{align}
Once $C_{re}(\rho_A) \le S_{vn}(\rho_{Adiag})$, it's possible to obtain an incomplete complementarity relation from this inequality:
\begin{equation}
    C_{re}(\rho_A) + P_{vn}(\rho_A) \le \ln d_A \label{eq:cr6},
\end{equation}
with $P_{vn}(\rho_A) := \ln d_A - S_{vn}(\rho_{Adiag}) = \ln d_A + \sum_{i = 0}^{d_A - 1} \rho^A_{ii} \ln \rho^A_{ii}$ being a measure of predictability, already defined in \cite{Englert, Maziero}. It is  possible to define this predictability measure once the diagonal elements of $\rho_A$ can be interpreted as a probability distribution, which is a consequence of the properties of $\rho_A$ \cite{Maziero}. The complementarity relation (\ref{eq:cr6}) is incomplete due the presence of correlations. However, $\rho_A$ is the subsystem of a bipartite pure quantum system $\ket{\Psi}_{A,B}$, which allow us to take $S_{vn}(\rho_A)$ as a measure of entanglement  of the subsystem A with B \cite{Vedral}. So, it's possible to interpret Eq. (\ref{eq:cre}) as a complete complementarity relation
\begin{equation}
         C_{re}(\rho_A) + P_{vn}(\rho_A) + S_{vn}(\rho_A) = \ln d_A \label{eq:cre_}.
\end{equation}

\section{Complementarity Relations for Tripartite Pure Quantum Systems }
\label{sec:tri}
Following the same logic as in the previous section, a tripartite pure quantum system can be represented by $\ket{\Psi}_{A,B,C} \in \mathcal{H}_{A} \otimes \mathcal{H}_{B} \otimes \mathcal{H}_C$. Let $\{\ket{i}_A\}_{i = 0}^{d_A - 1}$, $\{\ket{j}_B\}_{j = 0}^{d_B - 1}$,  $\{\ket{k}_C\}_{k = 0}^{d_C - 1}$ be the local orthonormal basis for the Hilbert spaces $\mathcal{H}_{A}$, $\mathcal{H}_{B}, \mathcal{H}_{C}$, respectively, so that
\begin{align}
    \rho_{A,B,C} = \ket{\Psi}_{A,B,C}\bra{\Psi} = \sum_{i,l=0}^{d_A - 1}\sum_{j,m=0}^{d_B - 1}\sum_{k,n=0}^{d_C -1}\rho_{ijk,lmn} \ket{i,j,k}_{A,B,C}\bra{l,m,n}
\end{align}
represents a tripartite pure quantum system. The subsystem A, for example, is represented by the reduced density operator
\begin{align}
    \rho_A = \sum_{i,l=0}^{d_A - 1} \rho^A_{il} \ket{i}_A\bra{l} = \sum_{i,l=0}^{d_A - 1}\sum_{j = 0}^{d_B - 1}\sum_{k = 0}^{d_C - 1}\rho_{ijk,ljk} \ket{i}_A\bra{l},
\end{align}
and similarly for the other subsystems. In general the state of the subsystem A is mixed, so, by exploring the properties of the reduced density matrix of A, one obtains an incomplete complementarity relation, just as before. Once the information content of A is being shared with B and C, it is natural to explore the purity of $\rho_{A,B,C}$:
\begin{equation}
   1 - \Tr \rho_{A,B,C}^2= 1 - \Big(\sum_{\overset{i = l}{\overset{j = m}{k = n}}} + \sum_{\overset{i \neq l}{\overset{j \neq m}{k \neq n}}} + \sum_{\overset{i = l}{\overset{j \neq m}{k \neq n}}} + \sum_{\overset{i \neq l}{\overset{j = m}{k \neq n}}} + \sum_{\overset{i \neq l}{\overset{j \neq m}{k = n}}} + \sum_{\overset{i = l}{\overset{j = m}{k \neq n}}} + \sum_{\overset{i = l}{\overset{j \neq m}{k = n}}} + \sum_{\overset{i \neq l}{\overset{j = m}{k = n}}}\Big)\abs{\rho_{ijk,lmn}}^2 = 0.
\end{equation}
This equation can be recast as 
\begin{align}
    P_{hs}(\rho_A) + C_{hs}(\rho_A) + C_{hs}^{nl}(\rho_{A|BC}) = \frac{d_A - 1}{d_A} \label{eq:cr4},
\end{align}
with $P_{hs}(\rho_A) = \sum_{i=0}^{d_A - 1} (\rho^A_{ii})^2 - 1/d_A = \sum_{i=0}^{d_A - 1}(\sum_{j = 0}^{d_B - 1}\sum_{k = 0}^{d_C - 1}\rho_{ijk,ijk})^2  - 1/d_A$, $C_{hs}(\rho_A) =  \sum_{i \neq k = 0}^{d_A - 1}\abs{\rho^A_{ik}}^2 = \sum_{i \neq l = 0}^{d_A - 1} \abs{\sum_{j = 0}^{d_B - 1}\sum_{k = 0}^{d_C - 1}\rho_{ijk,ljk}}^2 $, and $C_{hs}^{nl}(\rho_{A|BC})$ is the non-local coherence of A, shared with B and C, defined as
\begin{align}
    C_{hs}^{nl}(\rho_{A|BC}) := \sum_{i \neq l}\Big(\sum_{\overset{j \neq m}{k \neq n}} + \sum_{\overset{j = m}{k \neq n}} + \sum_{\overset{j \neq m}{k = n}}\Big)\abs{\rho_{ijk,lmn}}^2 - 2 \sum_{i \neq l}(\sum_{\overset{j = m}{k < n}} + \sum_{\overset{j < m}{k = n}} + \sum_{\overset{j < m}{k \neq n}}) \mathbb{R}e( \rho_{ijk,ljk} \rho^*_{imn,lmn}).
\end{align}
As we show below, $C_{hs}^{nl}(\rho_{A|BC})$ is equal to the linear entropy of A. The maximum of $C_{hs}^{nl}(\rho_{A|BC})$ is $(d_A - 1)/d_A$. Indeed 
\begin{align}
    1 - \Tr \rho_A^2 & = 1 - \sum_{i,l}\abs{\sum_{j,k}\rho_{ijk,ljk}}^2  = 1 - \sum_{i,l} \sum_{j,k}\sum_{m,n}\rho_{ijk,ljk}\rho^*_{imn,lmn}\\
    & = \sum_{i,j,k}\rho_{ijk,ijk} - \Big(\sum_{i =l} + \sum_{i \neq l}\Big)\Big( \sum_{\overset{j = m}{k = n}} + \sum_{\overset{j = m}{k \neq n}} + \sum_{\overset{j \neq m}{k = n}} + \sum_{\overset{j \neq m}{k \neq n}}\Big)\rho_{ijk,ljk}\rho^*_{imn,lmn}\\
    & = \sum_{i,j,k}\rho_{ijk,ijk}(1 - \rho_{ijk,ijk}) -  \Big(\sum_{i =l} + \sum_{i \neq l}\Big)\Big( \sum_{\overset{j = m}{k \neq n}} + \sum_{\overset{j \neq m}{k = n}} + \sum_{\overset{j \neq m}{k \neq n}}\Big)\rho_{ijk,ljk}\rho^*_{imn,lmn} -\sum_{\overset{i \neq l}{\overset{j = m}{k = n}}}\rho_{ijk,ljk}\rho^*_{imn,lmn}.
\end{align}
We notice the following identities
\begin{align}
  & \sum_{i,j,k}\rho_{ijk,ijk}(1 - \rho_{ijk,ijk}) = \Big( \sum_{\overset{i \neq l}{\overset{j \neq m}{k \neq n}}} + \sum_{\overset{i = l}{\overset{j \neq m}{k \neq n}}} + \sum_{\overset{i \neq l}{\overset{j = m}{k \neq n}}} + \sum_{\overset{i \neq l}{\overset{j \neq m}{k = n}}} + \sum_{\overset{i = l}{\overset{j = m}{k \neq n}}} + \sum_{\overset{i = l}{\overset{j \neq m}{k = n}}} + \sum_{\overset{i \neq l}{\overset{j = m}{k = n}}}\Big)\rho_{ijk,ijk}\rho_{lmn,lmn}, \\
  & \sum_{i \neq l}\sum_{\overset{j = m}{k = n}}\rho_{ijk,ljk}\rho^*_{imn,lmn}  =  \sum_{i \neq l}\sum_{\overset{j = m}{k = n}}\rho_{ijk,ijk}\rho^*_{lmn,lmn},  
\end{align}
where in the first identity we just rewrote the product of the diagonal elements using the fact that $\Tr \rho_{A,B,C} = 1$, while in the second identity we explored the purity of $\rho_{A,B,C}$. So,
\begin{align}
    1 - \Tr \rho_A^2 & = \Big( \sum_{\overset{i \neq l}{\overset{j \neq m}{k \neq n}}} + \sum_{\overset{i \neq l}{\overset{j = m}{k \neq n}}} + \sum_{\overset{i \neq l}{\overset{j \neq m}{k = n}}}\Big)\rho_{ijk,ijk}\rho_{lmn,lmn} - \sum_{i \neq l}\Big(\sum_{\overset{j = m}{k \neq n}} + \sum_{\overset{j \neq m}{k = n}} + \sum_{\overset{j \neq m}{k \neq n}}\Big)\rho_{ijk,ljk}\rho^*_{imn,lmn} \\
    & = \Big( \sum_{\overset{i \neq l}{\overset{j \neq m}{k \neq n}}} + \sum_{\overset{i \neq l}{\overset{j = m}{k \neq n}}} + \sum_{\overset{i \neq l}{\overset{j \neq m}{k = n}}}\Big)\abs{\rho_{ijk,lmn}}^2 - 2 \sum_{i \neq l}\Big(\sum_{\overset{j = m}{k < n}} + \sum_{\overset{j < m}{k = n}} + \sum_{\overset{j < m}{k \neq n}}\Big)\mathbb{R}e(\rho_{ijk,ljk}\rho_{imn,lmn}^*)\\
    & = C_{hs}^{nl}(\rho_{A|BC}),
\end{align}
which also shows that $E_M = \sqrt{2C_{hs}^{nl}(\rho_{A|BC})}$, where $E_M$ is the generalized concurrence defined in \cite{Bhaskara} for the special case of a tripartite pure quantum system. Now, if the system $A$ is not correlated with $B$ or $C$, then $A$ must be pure, since the impurity of a system is attributed to the correlations with other systems. Hence, if $\rho_{A,B,C} = \rho_A \otimes \rho_{B,C}$, i.e., if $\rho_{A,B,C}$ is, at least, a bi-separable state, then $\rho_A$ is pure and  $C_{hs}^{nl}(\rho_{A|BC}) = 0$. In addition, it's easy to see that $C_{hs}^{nl}(\rho_{A|BC}) \ge 0$ for any tripartite pure state:
\begin{align}
     C_{hs}^{nl}(\rho_{A|BC}) & = \sum_{i \neq l}\Big(\sum_{\overset{j \neq m}{k \neq n}} + \sum_{\overset{j = m}{k \neq n}} + \sum_{\overset{j \neq m}{k = n}}\Big)\abs{\rho_{ijk,lmn}}^2 - 2 \sum_{i \neq l}(\sum_{\overset{j = m}{k < n}} + \sum_{\overset{j < m}{k = n}} + \sum_{\overset{j < m}{k \neq n}}) \mathbb{R}e( \rho_{ijk,ljk} \rho^*_{imn,lmn})\\
     & \ge \sum_{i \neq l}\Big(\sum_{\overset{j \neq m}{k \neq n}} + \sum_{\overset{j = m}{k \neq n}} + \sum_{\overset{j \neq m}{k = n}}\Big)(\abs{\rho_{ijk,lmn}}^2 - \abs{\rho_{ijk,ljk}}\abs{\rho^*_{imn,lmn}})\\
     & = \sum_{i \neq l}\Big(\sum_{\overset{j \neq m}{k \neq n}} + \sum_{\overset{j = m}{k \neq n}} + \sum_{\overset{j \neq m}{k = n}}\Big)(\rho_{ijk,ijk}\rho_{lmn,lmn} - \sqrt{\rho_{ijk,ijk}\rho_{ljk,ljk}}\sqrt{\rho_{imn,imn}\rho_{lmn,lmn}})\\
     & = \frac{1}{2}\sum_{i \neq l}\Big(\sum_{\overset{j \neq m}{k \neq n}} + \sum_{\overset{j = m}{k \neq n}} + \sum_{\overset{j \neq m}{k = n}}\Big)(\sqrt{\rho_{ijk,ijk}\rho_{lmn,lmn}} - \sqrt{\rho_{imn,imn}\rho_{ljk,ljk}})^2\\
     & \ge 0,
\end{align}
where it was used the fact that $\rho_{A,B,C}$ is pure, and, for dummy indices that were summed, one can write
\begin{equation}
    \rho_{ijk,ijk}\rho_{lmn,lmn} = \frac{1}{2}(\rho_{ijk,ijk}\rho_{lmn,lmn} + \rho_{imn,imn}\rho_{ljk,ljk}).
\end{equation}
Finally, complementarity relations like (\ref{eq:cr4}) can be obtained for the subsystems B and C by exploring the purity of $\rho_{A,B,C}$. Also, we'll see in the next section that, for some tripartite quantum states, $C_{hs}^{nl}(\rho_{A|BC}) $ is related to some correlation measures already defined in the literature.

\section{Complementarity Relations for Multipartite Pure Quantum Systems}
\label{sec:multi}
Generalizing the procedure presented in the last two sections, here we report a general framework for obtaining complete complementarity relations for a subsystem which belongs to an arbitrary multipartite pure quantum system; we just have to explore the purity of the multipartite quantum system. So, let us consider a $n$-quanton pure state described by $\ket{\Psi}_{A_1,...,A_n} \in \mathcal{H}_{1} \otimes ... \otimes \mathcal{H}_{n}$. By defining a local orthonormal basis for each subsystem $A_m$, $\{\ket{i_m}_{A_m}\}_{i_m = 0}^{d_m - 1}$, $m = 1,...,n$, the state of the multipartite quantum system can be written as
\begin{align}
    \rho_{A_1, ..., A_n} = \sum_{i_1,...,i_n} \sum_{j_1,...,j_n} \rho_{i_1 ... i_n,j_1...j_n}\ket{i_1,...,i_n}_{A_1,...,A_n}\bra{j_1,...,j_n}.
\end{align}
Without loss of generality, let's consider the state of the subsystem $A_1$, which is obtained by tracing over the rest of the subsystems,
\begin{align}
    \rho_{A_1} = \sum_{i_1,j_1}\rho_{i_1,j_1}^{A_1}\ket{i_1}_{A_1}\bra{j_1} = \sum_{i_1,j_1}\sum_{i_2,...,j_n}\rho_{i_1 i_2 ... i_n, j_1 i_2 ... i_n}\ket{i_1}_{A_1}\bra{j_1},
\end{align}
for which the Hilbert-Schmidt quantum coherence and the corresponding predictability measure are given by
\begin{align}
    & C_{hs}(\rho_{A_1}) = \sum_{i_1 \neq j_1}\abs{\rho_{i_1,j_1}^{A_1}}^2 = \sum_{i_1 \neq j_1}\abs{\sum_{i_2,...,i_n}\rho_{i_1 i_2 ... i_n, j_1 i_2 ... i_n}}^2,\\
    & P_{hs}(\rho_{A_1}) = \sum_{i_1}(\rho_{i_1,i_1}^{A_1})^2 - 1/d_{A_1} = \sum_{i_1}(\sum_{i_2,...,i_n}\rho_{i_1 i_2 ... i_n, i_1 i_2 ... i_n})^2 - 1/d_{A_1}.
\end{align}
From these equations, an incomplete complementarity relation, $P_{hs}(\rho_{A_1}) + C_{hs}(\rho_{A_1})  \le (d_{A_1} - 1)/d_{A_1}$, is obtained by exploring the mixture of $\rho_{A_1}$, i.e., $1 - \Tr \rho_{A_1}^2 \ge 0$. Now, since $\rho_{A_1,...,A_n}$ is a pure quantum system, then
\begin{align}
1 - \Tr \rho_{A_1,...,A_n}^2=1 - \Big(\sum_{(i_1,...,i_n) = (j_1,...,j_n)} + \sum_{(i_1,...,i_n) \neq (j_1,...,j_n)}\Big) \abs{\rho_{i_1 i_2 ... i_n, j_1 j_2 ... j_n}}^2 = 0 \label{eq:pur},    
\end{align}
where
\begin{align}
    \sum_{(i_1,...,i_n) \neq (j_1,...,j_n)} \equiv \sum_{\overset{i_1 \neq j_1}{\overset{i_2 = j_2}{\overset{\vdots}{i_n = j_n}}}} + \sum_{\overset{i_1 = j_1}{\overset{i_2 \neq j_2}{\overset{\vdots}{i_n = j_n}}}} + ... + \sum_{\overset{i_1 = j_1}{\overset{i_2 = j_2}{\overset{\vdots}{i_n \neq j_n}}}} + \sum_{\overset{i_1 \neq j_1}{\overset{i_2 \neq j_2}{\overset{\vdots}{i_n = j_n}}}} + ... + \sum_{\overset{i_1 \neq j_1}{\overset{i_2 = j_2}{\overset{\vdots}{i_n \neq j_n}}}} + ... + \sum_{\overset{i_1 \neq j_1}{\overset{i_2 \neq j_2}{\overset{\vdots}{i_n \neq j_n}}}}.
\end{align}
Now, the purity condition in Eq. (\ref{eq:pur}) can be rewritten as a complete complementarity relation
\begin{align}
    P_{hs}(\rho_{A_1}) + C_{hs}(\rho_{A_1}) + C^{nl}_{hs}(\rho_{A_1|A_2,...,A_n}) = \frac{d_{A_1} - 1}{d_{A_1}},
\end{align}
where the non-local quantum coherence of system $A_1$, shared with $A_2,...,A_n$, is defined as
\begin{align}
    C^{nl}_{hs}(\rho_{A_1|A_2,...,A_n}) := \sum_{i_1 \neq j_1} \sum_{(i_2,...,i_n) \neq (j_2,...,j_n)}\Big(\abs{\rho_{i_1 i_2 ... i_n, j_1 j_2 ... j_n}}^2 -  \rho_{i_1 i_2 ... i_n, j_1 i_2 ... i_n}\rho_{i_1 j_2 ... j_n, j_1 j_2 ... j_n}^*\Big).
\end{align}
To show that $E_M = \sqrt{2C^{nl}_{hs}(\rho_{A_1|A_2,...,A_n})}$, where $E_M$ is the generalized concurrence defined in \cite{Bhaskara}, and  $0 \le C^{nl}_{hs}(\rho_{A_1|A_2,...,A_n}) \le (d_{A_1} -1) /d_{A_1}$, where $C^{nl}_{hs}(\rho_{A_1|A_2,...,A_n}) = 0$ iff $\rho_{A_1,...,A_n} = \rho_{A_1}\otimes \rho_{A_2,...,A_n}$, and hence $\rho_{A_1}$ is pure, it's enough to notice that $C^{nl}_{hs}(\rho_{A_1|A_2,...,A_n})$ is equal to the linear entropy of $A_1$:
\begin{align}
    1 - \Tr(\rho_{A_1}^2)& = 1 - \sum_{i_1, j_1}\abs{\sum_{i_2,...,i_n}\rho_{i_1 i_2 ... i_n, j_1 i_2 ... i_n}}^2\\
    & = \sum_{i_1,...,i_n}\rho_{i_1 i_2 ... i_n, i_1 i_2 ... i_n}
    - \Big(\sum_{i_1 = j_1} + \sum_{i_1 \neq j_1}\Big)\sum_{i_2,...,i_n}\sum_{j_2,...,j_n}\rho_{i_1 i_2 ... i_n, j_1 i_2 ... i_n}\rho^*_{i_1 j_2 ... j_n, j_1 j_2 ... j_n}\\
    & = \sum_{(i_1,...,i_n) \neq (j_1,...,j_n)}\abs{\rho_{i_1 i_2 ... i_n, j_1 j_2 ... j_n}}^2 - \Big(\sum_{i_1 = j_1} + \sum_{i_1 \neq j_1}\Big)\sum_{(i_2,...,i_n) \neq (j_2,...,j_n)}\rho_{i_1 i_2 ... i_n, j_1 i_2 ... i_n}\rho^*_{i_1 j_2 ... j_n, j_1 j_2 ... j_n}\\ & - \sum_{i_1 \neq j_1}\sum_{(i_2,...,i_n) = (j_2,...,j_n)}\rho_{i_1 i_2 ... i_n, j_1 i_2 ... i_n}\rho^*_{i_1 j_2 ... j_n, j_1 j_2 ... j_n} \nonumber \\
    & = \sum_{i_1 \neq j_1} \sum_{(i_2,...,i_n) \neq (j_2,...,j_n)}\Big(\abs{\rho_{i_1 i_2 ... i_n, j_1 j_2 ... j_n}}^2 - \rho_{i_1 i_2 ... i_n, j_1 i_2 ... i_n}\rho_{i_1 j_2 ... j_n, j_1 j_2 ... j_n}^*\Big)\\
    & = C^{nl}_{hs}(\rho_{A_1|A_2,...,A_n}).
\end{align}

Lastly, we would like the emphasize the main idea of our work: Due to the purification theorem \cite{Mark}, it is always possible to consider a multipartite pure system, and then explore its purity to obtain complete complementarity relations for any of its subsystems. On the other hand, if we consider $n$-partite mixed quantum systems described by $\rho_{A_1,...,A_n}$, then we have two possibilities:
\begin{enumerate}
    \item By exploring $1 - \Tr \rho^2_{A_1,...,A_n} \ge 0$, the equality in $C_{hs} + P_{hs} + C_{hs}^{nl} = (d_{A_1} -1)/d_{A_1}$ becomes an inequality, and $C_{hs}^{nl} \neq 1 - \Tr \rho^2_{A_1}$ with $C_{hs}^{nl}$ measuring only quantum correlations of $A_1$ with the rest of the system. 
    
    \item $S_l(\rho_{A_1}) = 1 - \Tr \rho^2_{A_1}$ measures not only quantum correlations, but quantifies, more generally, the noise introduced by interactions with the environment. Then, we can interpret the entropy functional $S_l(\rho_{A_1})$ as the mixedness of the subsystem $A_1$, as defined in \cite{Dhar}, and we'll still have a complete complementarity relation (CCR), even though it is not possible to derive such CCR directly from our framework. The same interpretation can be given for Eq. (\ref{eq:cre_}).
\end{enumerate}

\section{Examples \& Relations between Quantum Correlations}
\label{sec:rqc}
\subsection{Bipartite states}
According to \cite{Tan}, with respect to the local reference bases $\{\ket{i}_A\}_{i = 0}^{d_A - 1}$, $\{\ket{j}_B\}_{j = 0}^{d_B - 1}$, the correlated coherence for a bipartite quantum system is given by $C^c(\rho_{A,B}) := C(\rho_{A,B}) - C(\rho_{A}) - C(\rho_{B})$ for some measure of coherence. As showed by the same authors, for the $l_1$-norm $C^c_{l_1}(\rho_{A,B}) \ge 0$ for any bipartite quantum state. Also, it's worth pointing out that $C^c_{l_1}(\rho_{A,B}) = 0$ for $\rho_{A,B} = \rho_A \otimes \sigma_B$, where $\sigma_B$ is an incoherent state. However, for separable uncorrelated states, $\rho_{A,B} = \rho_A \otimes \rho_B = \sum_{i,k = 0}^{d_A -1}\sum_{j,l = 0}^{d_B -1} \rho_{i,k}^A \rho_{j,l}^B \ket{i,j}_{A,B}\bra{k,l} $,
\begin{align}
    C^c_{l_1}(\rho_{A}\otimes \rho_B) & = \Big( \sum_{\overset{i \neq k}{j \neq l}} + \sum_{\overset{i \neq k}{j = l}} + \sum_{\overset{i = k}{j \neq l}}\Big)\abs{\rho_{ik}^A}\abs{\rho_{jl}^B} - \sum_{i\neq k}\abs{\rho_{ik}^A} - \sum_{j\neq l}\abs{\rho_{jl}^A}\\
    & = \sum_{\overset{i \neq k}{j \neq l}}\abs{\rho_{ik}^A}\abs{\rho_{jl}^B} \ge 0, \\
\end{align}
which implies that $C_{l_{1}}(\rho_{A}\otimes \rho_B) \neq C_{l_{1}}(\rho_{A}) + C_{l_{1}}(\rho_{B})$, while for the relative entropy of coherence the equality is always satisfied, i.e., $C_{re}^{c}(\rho_{A}\otimes \rho_B)=0$ \cite{Kraft}. For the Hilbert-Schmidt measure of coherence, it turns out that $C_{hs}^c(\rho_{A,B})$ can be negative for some cases, as for example for a bipartite quantum system such that $\rho_{A,B} = \rho_A \otimes \sigma_B$, where $\sigma_B$ is an incoherent state. Besides that, if the off-diagonal elements of that reduced density matrices satisfy the following properties
\begin{align}
    & \abs{\rho^A_{ik}}^2 = \abs{\sum_j \rho_{ij,kj}}^2 = \sum_j\abs{\rho_{ij,kj}}^2, \ \forall i \neq k \label{eq:cd1},\\
    & \abs{\rho^B_{jl}}^2 = \abs{\sum_i \rho_{ij,il}}^2 = \sum_i\abs{\rho_{ij,il}}^2, \label{eq:cd2} \ \forall j \neq l,
\end{align}
it is straightforward to show that $C_{hs}^c(\rho_{A,B}) \ge 0$. For multipartite states, this will remain true if the reduced density matrices of the subsystems satisfy properties similar to (\ref{eq:cd1}) and (\ref{eq:cd2}). All the states considered in this article, except the last one, satisfies these conditions.\\

Now, for a bipartite quantum system in the state $\ket{\Psi}_{A,B} = x\ket{0,1}_{A,B} + \sqrt{1 - x^2}\ket{1,0}_{A,B}$, with $x \in [0,1]$, we have
\begin{align}
    & C_{l_1}^c(\rho_{A,B}) = C^{(2)}_{A,B} = 2x\sqrt{1 - x^2}, \label{ex1i} \\
    & P_{l_1}(\rho_A) = P_{l_1}(\rho_B) = 1 - 2x\sqrt{1 - x^2},\\
    & C_{hs}^{nl}(\rho_{A|B}) = C^c_{hs}(\rho_{A,B}) = \frac{1}{2}[C^{(2)}_{A,B}]^2 = 2x^2(1 - x^2), \\
    & P_{hs}(\rho_A) = P_{hs}(\rho_B) = \frac{1}{2} P^2_A = \frac{1}{2} P^2_B = 1/2 - 2x^2(1 - x^2),\\
    & S_{vn}(\rho_A) = S_{vn}(\rho_B) = -x^2 \ln x^2 - (1 - x^2) \ln (1 - x^2), \\    
    & P_{vn}(\rho_A) = P_{vn}(\rho_B) = \ln 2 + x^2 \ln x^2 + (1 - x^2) \ln (1 - x^2), \label{ex1f}
\end{align}
\begin{figure}[t]
\includegraphics[scale=0.6]{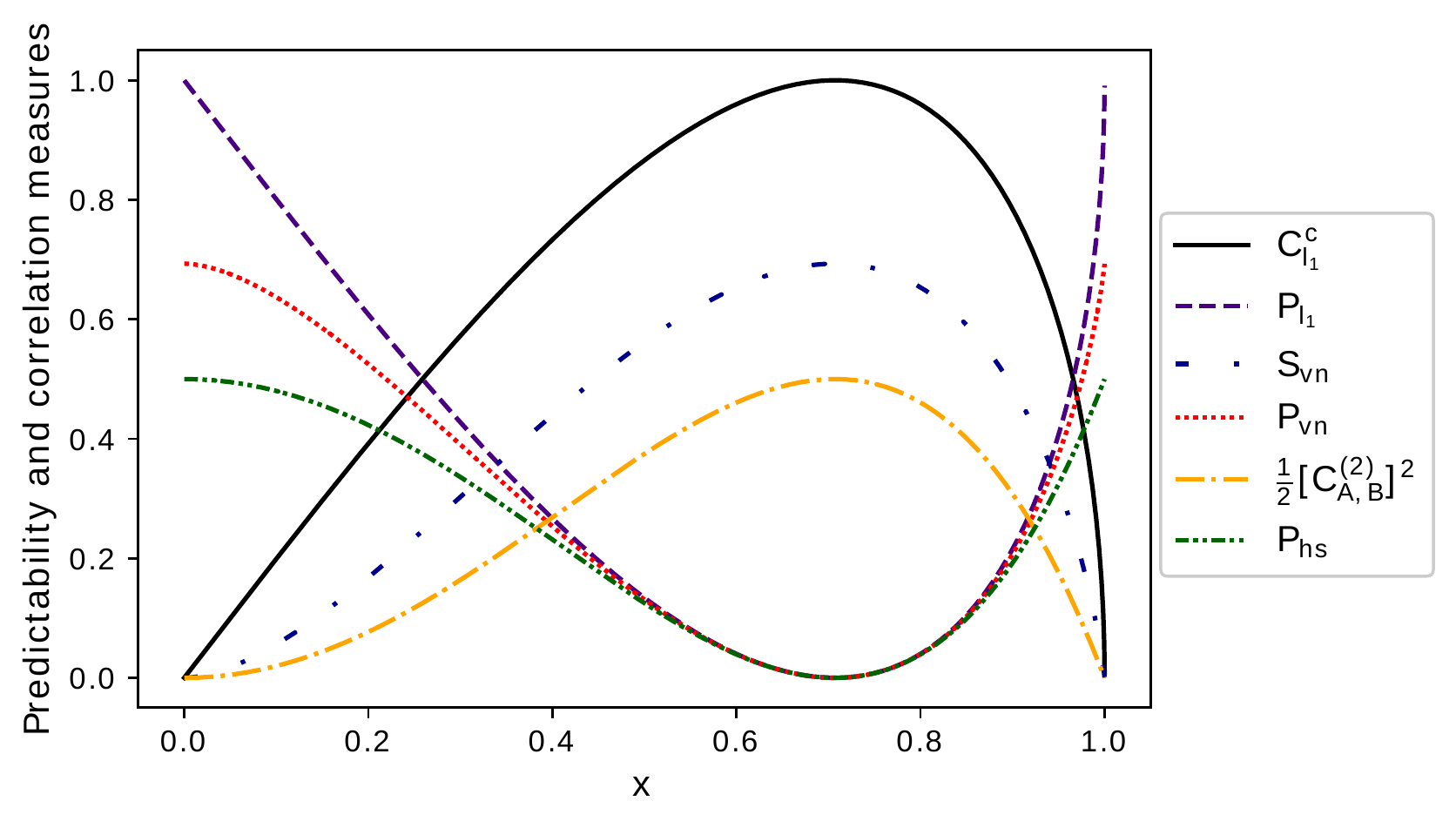}
\caption{Comparison between different measures of predictability and their respectively correlations measures of Eqs. (\ref{ex1i})-(\ref{ex1f}) as a function of $x$.}
\label{fig:mes}
\end{figure}
where $C^{(2)}_{A,B}$ is the concurrence measure for entanglement \cite{Jakob, Bergou}, and $P_{l_1}(\rho) := d - 1 - \sum_{j \neq k}\sqrt{\rho_{jj}\rho_{kk}}$ is a measure of predictability obtained in \cite{Maziero}. One can easily see that $C^{(2)}_{A,B} = C^c_{l_1}(\rho_{A,B})$. In Fig. \ref{fig:mes}, we plotted the different measures of predictability and correlation for comparison. As expected, the measures of predictability (and correlation) reach their maximum and minimum values in the same point of the domain, although the maximum and minimum values of the functions differ from one measure to the other. However, the inflection point in the domain is not the same among the measures of predictability (and correlation), which is interesting once the inflection point represents the point in the domain where the predictability and the respective measure of correlation reach the same value. Lastly, we observe that these measures are all equivalent and all have the same physical significance, since they all meet the criteria established by the literature \cite{Durr, Englert}. This, of course, can change if an experiment or a physical situation that distinguishes them appears. Then it will be necessary to modify or add some criteria to exclude some of the measures. Besides that, for each coherence measure used to quantify the wave aspect of the quantum, there's a corresponding predictability measure, since both have to reach the same maximum value. For instance, $C_{hs}, P_{hs}$ reaches the maximum value given by $(d - 1)/d$, whereas $C_{re}, P_{vn}$ reaches the maximum value given by $\ln d$. Therefore, it is necessary to consider different correlation measures to complete each of the complementarity relations. 

\begin{figure}[t]
\includegraphics[scale=0.6]{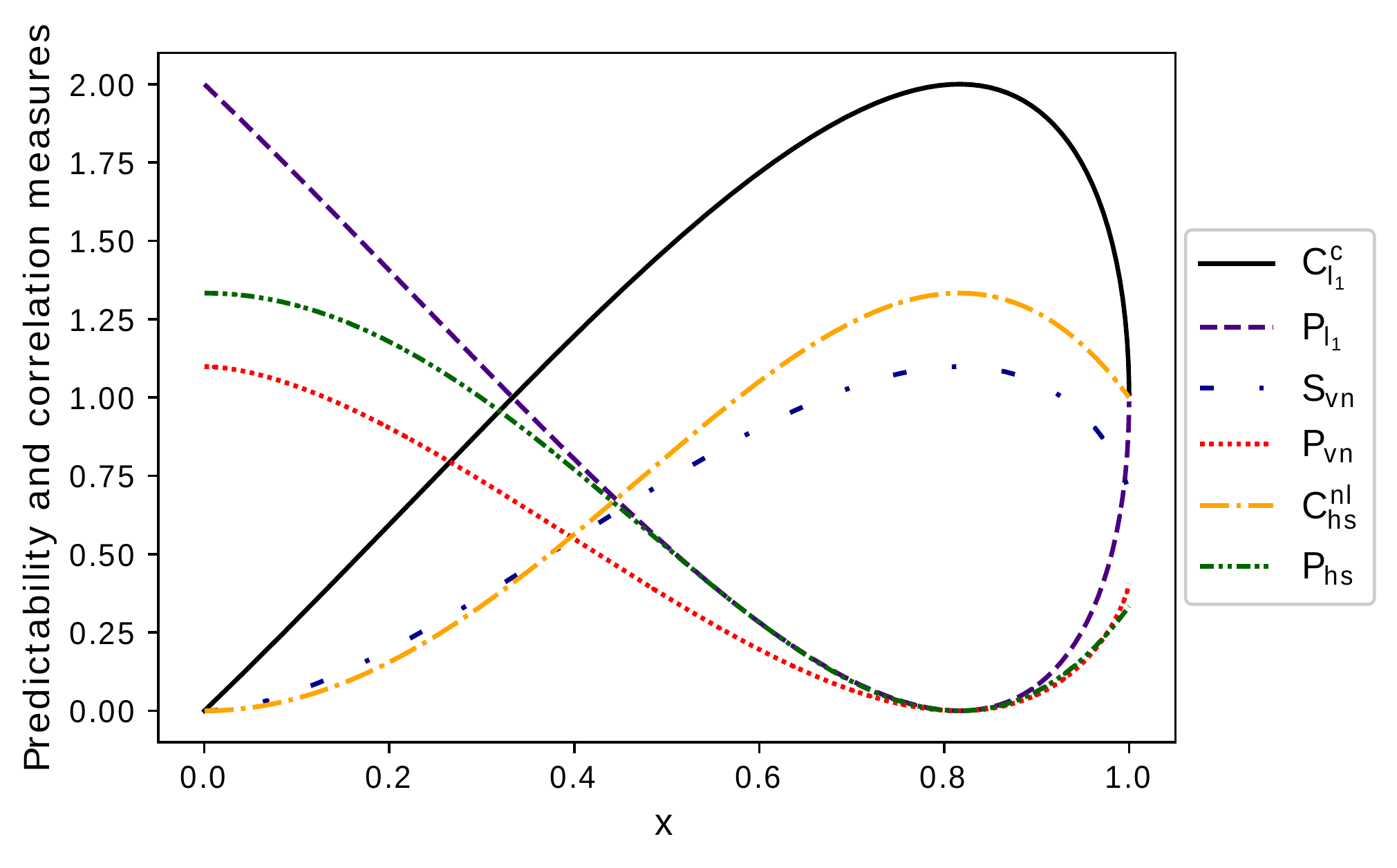}
\caption{Comparison between different measures of predictability and their respective correlation measures of the state (\ref{ex2}) as a function of $x$.}
\label{fig:me}
\end{figure}

Another example was given by Jacob and Bergou in \cite{Bergou}, where they considered the following state:
\begin{align}
    \ket{\Psi}_{A,B} = \frac{x}{\sqrt{2}}\ket{0,0}_{A,B} + \frac{x}{\sqrt{2}}\ket{1,1}_{A,B} + \sqrt{1 - x^2} \ket{2,2}_{A,B}, \label{ex2}
\end{align}
with $x \in [0,1]$. The predictability and concurrence measures defined in \cite{Bergou} are given by $P_A^2= P_B^2 = 3x^4 - 4x^2 + 4/3$ and $[C^{(3)}_{A,B}]^2 = 4x^2 - 3x^4$, respectively, so that $P_A^2 + [C^{(3)}_{A,B}]^2 = P_B^2 + [C^{(3)}_{A,B}]^2 = 4/3$. However, if we change the measure of predictability, the measure of correlation also has to be changed in order to obtain a complementarity relation that saturates. For $P_{l_1}(\rho_A) = P_{l_1}(\rho_B) = 2(1 - x^2/2 - \sqrt{2x^2(1 - x^2)})$, the corresponding correlation measure is the $l_1$-norm correlated coherence $C^c_{l_1}(\rho_{A,B}) = 2(x^2/2 + \sqrt{2x^2(1 - x^2)})$, thus $P_{l_1}(\rho_A) + C^c_{l_1}(\rho_{A,B}) = P_{l_1}(\rho_B) + C^c_{l_1}(\rho_{A,B}) = 2$. An interesting fact is that if we use the non-normalized predictability measure defined by Roy and Qureshi in \cite{Qureshi}, it is not possible to obtain a complete complementarity relation, neither using the correlated coherence nor with the concurrence. In addition, by writing the state as $\ket{\Psi}_{A,B} = a_{00}\ket{0,0}_{A,B} + a_{11}\ket{1,1}_{A,B} + a_{22} \ket{2,2}_{A,B}$ such that $a_{00} = a_{11} = x/\sqrt{2}$ and $a_{22} = \sqrt{1 - x^2}$, one can see that the off-diagonal elements of the density matrix related to $C^c_{l_1}(\rho_{A,B})$ and to $C^{(3)}_{A,B}$ are the same, the only difference between these measures is the form of the function: $C^c_{l_1}(\rho_{A,B}) = 2(\abs{\rho_{00,11}} + \abs{\rho_{00,22}} + \abs{\rho_{11,22}})$, while $C^{(3)}_{A,B} = 2 \sqrt{\rho_{00,11}^2 + \rho_{00,22}^2 + \rho_{11,22}^2}$. In this case, one correlation measure can't be written as a function of the other correlation measure. In Fig. \ref{fig:me}, we plotted different measures of predictability and correlation for comparison, including $P_{vn}$ and $S_{vn}$ defined in Sec. \ref{sec:bi}. The analysis is the same as before. 

\subsection{Tripartite states}

\begin{figure}[t]
\includegraphics[scale=0.65]{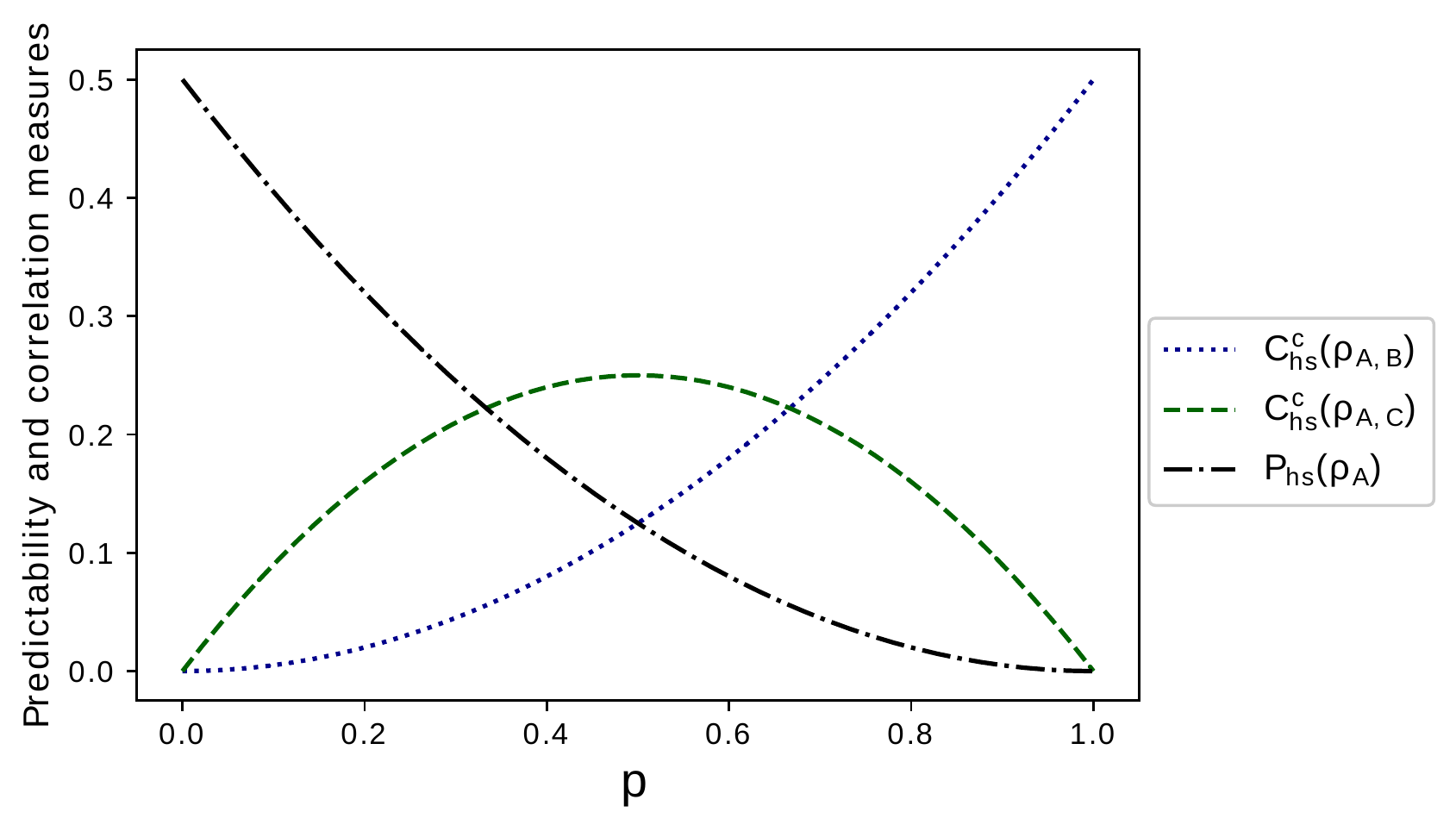}
\caption{\normalsize Predictability and correlated coherence measures for the subsystem A of the W state as a function of $p$.}
\label{fig:mes_}
\end{figure}

We start by considering as examples the GHZ and the W states \cite{Vidal}. For the GHZ state, $\ket{GHZ} = a_{000} \ket{0,0,0}_{A,B,C} + a_{111} \ket{1,1,1}_{A,B,C}$ with $\abs{a_{000}}^2 + \abs{a_{111}}^2 = 1$, we have $C_{hs}^{nl}(\rho_{A|BC}) = C_{hs}^{nl}(\rho_{B|AC}) = C_{hs}^{nl}(\rho_{C|AB}) = 2 \abs{a_{000}}^2 \abs{a_{111}}^2 = 2 \abs{a_{000}^2 a_{111}^2} = \frac{1}{2} \tau_3$, where $\tau_3$ is the tangle measure for entanglement \cite{Coffman, Uhlmann}. So the GHZ state satisfies the complementarity relations obtained in Sec. \ref{sec:tri}. For the W state, $\ket{W}_{A,B,C} = \sqrt{1 - p}\ket{0,0,1}_{A,B,C} + \sqrt{p/2}\ket{0,1,0}_{A,B,C} + \sqrt{p/2}\ket{1,0,0}_{A,B,C}$ with $p \in [0,1]$, we have $C_{hs}^{nl}(\rho_{A|BC}) = C_{hs}^c(\rho_{A,B}) + C_{hs}^c(\rho_{A,C}) = p^2/2 + p(1 - p)$, where $\rho_{A,B}, \rho_{A,C}$ are the reduced density matrices of $\ket{W}$. Once $P_{hs}(\rho_{A}) = 1/2 - p + p^2/2$, we can write the following complementarity relation for the subsystem A:
\begin{align}
    P_{hs}(\rho_A) + C_{hs}^{nl}(\rho_{A|BC}) = P_{hs}(\rho_A) + C_{hs}^c(\rho_{A,B}) + C_{hs}^c(\rho_{A,C}) = \frac{1}{2},
\end{align}
and similarly for the other subsystems. The functions appearing in the last equation are plotted in Fig. \ref{fig:mes_}. Thus, one can see that the complementarity relation for a subsystem is completed by its bipartite correlations with the other subsystems. We also notice that in this case the concurrence measure, for the reduced density matrices $\rho_{A,B}, \rho_{A,C}$, and $\rho_{B,C}$, is related to the Hilbert-Schmidt correlated coherences. Also, if we use the predictability measure $P_{l_1}(\rho_A)$, defined in \cite{Maziero}, it is not possible to obtain a complete complementarity relation using the $l_1$-norm correlated coherence.

Following, we regard the tripartite state $\ket{\Psi}_{A,B,C} = \lambda_1 \ket{0,0,0}_{A,B,C} + \lambda_2 \ket{0,0,1}_{A,B,C} + \lambda_3\ket{0,1,0}_{A,B,C} + \lambda_4\ket{1,0,0}_{A,B,C} + \lambda_5 \ket{1,1,1}_{A,B,C}$ with $\abs{\lambda_1}^2 + \abs{\lambda_2}^2 + \abs{\lambda_3}^2 + \abs{\lambda_4}^2 + \abs{\lambda_5}^2= 1$. The tangle quantifies only the correlations shared by the three subsystems simultaneously. Since the subsystems A, B and C satisfy conditions similar to (\ref{eq:cd1}) and (\ref{eq:cd2}), we can easily see that
\begin{align}
    C_{hs}^{nl}(\rho_{A|BC}) & = C_{hs}^{c}(\rho_{A,B,C}) - C_{hs}^{c}(\rho_{B,C}) \\ & = 2(\abs{\lambda_1}^2\abs{\lambda_5}^2 + \abs{\lambda_2}^2\abs{\lambda_4}^2 + \abs{\lambda_2}^2\abs{\lambda_5}^2 + \abs{\lambda_3}^2\abs{\lambda_4}^2 + \abs{\lambda_3}^2\abs{\lambda_5}^2).
\end{align}
Once $ P_{hs}(\rho_A) = (\abs{\lambda_1}^2 + \abs{\lambda_2}^2 + \abs{\lambda_3}^2)^2 + (\abs{\lambda_4}^2 + \abs{\lambda_5}^2)^2 - 1/2$, and $C_{hs}(\rho_A) = 2 \abs{\lambda_1}^2 \abs{\lambda_4}^2$, we have 
\begin{align}
    P_{hs}(\rho_A) + C_{hs}(\rho_A) + C_{hs}^{nl}(\rho_{A|BC}) = P_{hs}(\rho_A) + C_{hs}(\rho_A) + C_{hs}^{c}(\rho_{A,B,C}) - C_{hs}^{c}(\rho_{B,C}) = \frac{1}{2} \label{eq:cr5}.
\end{align}
As the correlated coherence of a tripartite system measures the coherence shared by the three subsystems simultaneously, and the coherence shared between two subsystems, we can see that the coherence shared between B and C is irrelevant for the complementarity relation of the subsystem A. It's also worth pointing out that for the others predictability measures mentioned before \cite{Qureshi, Maziero}, it is not possible to obtain a complete complementarity relation analogous to (\ref{eq:cr5}).

Finally, let's consider the following state $\ket{\Psi}_{A,B,C} = \lambda_1 \ket{0,0,0}_{A,B,C} + \lambda_2\ket{0,1,1}_{A,B,C} + \lambda_3 \ket{1,0,0}_{A,B,C} + \lambda_4 \ket{1,1,1}_{A,B,C}$, with $\abs{\lambda_1}^2 + \abs{\lambda_2}^2 + \abs{\lambda_3}^2 + \abs{\lambda_4}^2 = 1$ \cite{Acin}. In this case $C_{hs}^{nl}(\rho_{A|BC})  \neq C_{hs}^{c}(\rho_{A,B,C}) - C_{hs}^{c}(\rho_{B,C})$, since the reduced density matrix $\rho_A$ doesn't satisfy conditions similar to (\ref{eq:cd1}) and (\ref{eq:cd2}). In a straightforward calculation, one can show that
\begin{align}
    & C_{hs}^{nl}(\rho_{A|BC}) = 2 (\abs{\lambda_1}^2 \abs{\lambda_4}^2 + \abs{\lambda_2}^2 \abs{\lambda_3}^2 - 2 \mathbb{R}e(\lambda_1 \lambda_2^* \lambda_3^* \lambda_4)),\\
    & C_{hs}(\rho_A) = 2\abs{\lambda_1 \lambda_3^* + \lambda_2 \lambda_4^*}^2,\\
    & P_{hs}(\rho_A) = (\abs{\lambda_1}^2 + \abs{\lambda_2}^2)^2 + (\abs{\lambda_3}^2 + \abs{\lambda_4}^2)^2- 1/2.
\end{align}
Hence $ P_{hs}(\rho_A) + C_{hs}(\rho_A) + C_{hs}^{nl}(\rho_{A|BC}) = 1/2$. It's interesting to notice that the bipartite reduced density matrices of $\ket{\Psi}_{A,B,C}$ are given by
\begin{align}
     \rho_{A,B} & = \Big(\abs{\lambda_1}^2 \ketbra{0} + (\lambda_1 \lambda_3^* \ketbra{0}{1} + t.c.) + \abs{\lambda_3}^2 \ketbra{1}\Big)\otimes \ketbra{0} \\ & + \Big(\abs{\lambda_2}^2 \ketbra{0} + (\lambda_2 \lambda_4^* \ketbra{0}{1} + t.c.) + \abs{\lambda_4}^2 \ketbra{1}\Big)\otimes \ketbra{1},  \nonumber \\
    \rho_{A,C} & = \Big(\abs{\lambda_1}^2 \ketbra{0} + (\lambda_1 \lambda_3^* \ketbra{0}{1} + t.c.) + \abs{\lambda_3}^2 \ketbra{1}\Big)\otimes \ketbra{0} \\ &+ \Big(\abs{\lambda_2}^2 \ketbra{0} + (\lambda_2 \lambda_4^* \ketbra{0}{1} + t.c.) + \abs{\lambda_4}^2 \ketbra{1}\Big)\otimes \ketbra{1},  \nonumber  \\
    \rho_{B,C} & = \Big(\abs{\lambda_1}^2 + \abs{\lambda_3}^2\Big)\ketbra{0,0} + \Big(\lambda_1 \lambda_2^* + \lambda_3 \lambda_4^*\Big) \ketbra{0,1}{1,0} + t.c. + \Big(\abs{\lambda_2}^2 + \abs{\lambda_4}^2\Big)\ketbra{1,1}, 
\end{align}
where $t.c.$ stands for the transpose conjugate. One can see that $\rho_{A,B}$, and $\rho_{A,C}$ are quantum-classical states \cite{Adesso}. Meanwhile, $\rho_{B,C}$ is an entangled state once its reduced density matrices $\rho_B$ and $\rho_C$ are mixed/incoherent states. On the other hand, by writing  
\begin{equation}
\ket{\Psi}_{A,B,C} = (\lambda_1 \ket{0}_A + \lambda_3 \ket{1}_A) \otimes \ket{0,0}_{B,C} + (\lambda_2 \ket{0}_A + \lambda_4 \ket{1}_A) \otimes \ket{1,1}_{B,C},    
\end{equation}
one can expect that the correlations between the subsystem $A$ and the subsystem $BC$, where $BC$ is taken as one system, is due to entanglement. But, when analyzing the subsystem $BC$ separately, the correlations between $A$ and $B$ (or $C$) is due to quantum discord.

\section{Conclusions}
\label{sec:conc}
 By exploring the purity of multipartite pure quantum states, we showed that it is possible to obtain complementarity relations that characterize completely a quantum system. For the bipartite case, the complete complementarity relation (CCR) obtained is equivalent to the complementarity relation proposed by Jakob and Bergou \cite{Jakob, Bergou}. Also, in the same framework, it was possible to obtain a new complementarity relation, just by reinterpreting the definition of the relative entropy of coherence. For the tripartite case, we obtained a CCR with the correlation measure being equivalent to the generalized concurrence obtained in \cite{Bhaskara}. The procedure of obtaining CCR exploring the purity of the density matrix allowed us to create a general framework to derive such complementarity relations, and enabled us to generalize the work of Jakob and Bergou for the multipartite case. Our approach can be seen as natural, since we derived the CCRs directly from the hypothesis that the multipartite quantum system is pure, finding the complementarity measures within the expression $1 - \Tr \rho^2_{A_1,..., A_n} = 0$. Besides, through simple examples, we showed that if one changes the predictability measure, one has also to change the correlation measure in order to obtain a CCR for pure states. As quoted by Bohr \cite{Wit}: \textit{"... evidence  obtained  under different experimental conditions  cannot  be  comprehended  within  a single  picture,  but  must  be  regarded  as  complementary in  the  sense  that  only  the  totality  of  the  phenomena exhausts  the  possible  information  about  the  objects"}. However, to fully characterize a quanton, it is not enough to consider its wave-particle aspect, one also has to regard its correlations with other systems. 
 
\begin{acknowledgments}
This work was supported by the Coordena\c{c}\~ao de Aperfei\c{c}oamento de Pessoal de N\'ivel Superior (CAPES), process 88882.427924/2019-01, and by the Instituto Nacional de Ci\^encia e Tecnologia de Informa\c{c}\~ao Qu\^antica (INCT-IQ), process 465469/2014-0.
\end{acknowledgments}


\begin{thebibliography}{10}
\bibliographystyle{apsrev4-1}
\bibitem{Bohr} N. Bohr, The quantum postulate and the recent development of atomic theory, Nature 121, 580 (1928).
\bibitem{Leblond} According with J. -M. L\'evy-Leblond, the term "quanton" was given by M. Bunge. The usefulness of this term is that one can refer to a generic quantum system without using words like particle or wave: J.-M. L\'evy-Leblond, On the Nature of Quantons, Science and Education 12, 495 (2003).
\bibitem{Wootters} W. K. Wootters, W. H. Zurek,  Complementarity in the double-slit experiment: Quantum nonseparability and a quantitative statement of Bohr's principle, Phys. Rev. D 19, 473 (1979).
\bibitem{Engle} B.-G. Englert, Fringe Visibility and Which-Way Information: An Inequality, Phys. Rev. Lett. 77, 2154 (1996).
\bibitem{Yasin} D. M. Greenberger, A. Yasin, Simultaneous wave and particle knowledge in a neutron interferometer, Phys. Lett. A 128, 391 (1988).
\bibitem{Jaeger} G. Jaeger, A. Shimony, L. Vaidman, Two interferometric complementarities, Phys. Rev. A 51, 54 (1995).
\bibitem{Durr} S. D\"urr, Quantitative wave-particle duality in multibeam interferometers, Phys. Rev. A 64, 042113 (2001).
\bibitem{Englert} B.-G. Englert, D. Kaszlikowski, L. C. Kwek, W. H. Chee, Wave-particle duality in multi-path interferometers: General concepts and three-path interferometers, Int. J. Quant. Inf. 6, 129 (2008).
\bibitem{Baumgratz} T. Baumgratz, M. Cramer,  M. B. Plenio, Quantifying Coherence, Phys. Rev. Lett. 113, 140401 (2014).
\bibitem{Bera} M. N. Bera, T. Qureshi, M. A. Siddiqui, A. K. Pati, Duality of Quantum coherence and path distinguishability, Phys. Rev. A 92, 012118 (2015).
\bibitem{Bagan} E. Bagan, J. A. Bergou, S. S. Cottrell, M. Hillery, Relations between coherence and path information, Phys. Rev. Lett. 116, 160406 (2016).
\bibitem{Tabish} T. Qureshi,  Coherence, interference and visibility, Quanta 8, 24 (2019).
\bibitem{Mishra} S. Mishra, A. Venugopalan, T. Qureshi, Decoherence and visibility enhancement in multi-path interference, Phys. Rev. A 100, 042122 (2019)
\bibitem{Xu} B.-M. Xu, Z. C. Tu, J. Zou,  Duality in quantum work, Phys. Rev. A 101, 022113 (2020).
\bibitem{Vlatko} B. C. Hiesmayr, V. Vedral,  Thermodynamical versus optical complementarity, arXiv:quant-ph/0501015 (2005).
\bibitem{Angelo} R. M. Angelo, A. D. Ribeiro, Wave-particle duality: An information-based approach, Found. Phys. 45, 1407 (2015).
\bibitem{Coles} P. J. Coles, Entropic framework for wave-particle duality in multipath interferometers, Phys. Rev. A 93, 062111 (2016).
\bibitem{Hillery} E. Bagan, J. Calsamiglia,  J. A. Bergou, M. Hillery, Duality games and operational duality relations, Phys. Rev. Lett. 120, 050402 (2018).
\bibitem{Qureshi} P. Roy, T. Qureshi,  Path predictability and quantum coherence in multi-slit interference. Phys. Scr. 94, 095004 (2019),
\bibitem{Maziero} M. L. W. Basso, D. S. S. Chrysosthemos, J. Maziero,  Quantitative wave-particle duality relations from the density matrix properties, Quant. Inf. Process. 19, 254 (2020).
\bibitem{Janos} M. Jakob, J. A. Bergou, Quantitative complementarity relations in bipartite systems: Entanglement as a physical reality, Opt. Comm. 283, 827 (2010).
\bibitem{Bruss} D. Bruss, Characterizing entanglement, J.Math. Phys. 43, 4237 (2002).
\bibitem{Horodecki} R. Horodecki, P. Horodecki, M. Horodecki, K. Horodecki, Quantum entanglement, Rev. Mod. Phys. 81, 865 (2009).
\bibitem{Woot} W. K. Wootters, Entanglement of formation of an arbitrary state of two qubits. Phys. Rev. Lett. 80, 2245 (1998).
\bibitem{Qian} X.-F. Qian et al, Turning off quantum duality, Phys. Rev. Research 2, 012016 (2020).
\bibitem{Jakob} M. Jakob, J. A. Bergou,  Complementarity and entanglement in bipartite qudit systems, Phys. Rev. A 76, 052107 (2007).
\bibitem{Bergou} M. Jakob, J. A. Bergou, Generalized complementarity relations in composite quantum systems of arbitrary dimensions, Int. J. Mod. Phys. B 20, 1371 (2006).
\bibitem{Rungta} P. Rungta, V. Bu\^zek, C. M. Caves, M. Hillery, G. J. Milburn, Universal state inversion and concurrence in arbitrary dimensions, Phys. Rev. A 64, 042315 (2001).
\bibitem{Huber} B. C. Hiesmayr, M. Huber, Multipartite entanglement measure for all discrete systems, Phys. Rev. A 78, 012342 (2008).
\bibitem{Bromley}G. Adesso, T. R. Bromley, M. Cianciaruso, Measures and applications of quantum correlations, J. Phys. A: Math. Theor. 49, 473001 (2016).
\bibitem{Adesso} G. Adesso, M. Cianciaruso, T. R. Bromley, An introduction to quantum discord and non-classical correlations beyond entanglement, arXiv:1611.01959 (2016).
\bibitem{Xi} Z. Xi,  Y. Li, H. Fan, Quantum coherence and correlations in quantum system, Sci. Rep. 5, 10922 (2015).
\bibitem{Xiz} Z. Xi, Coherence distribution in multipartite systems, J. Phys. A: Math. Theor. 51, 414016 (2018).
\bibitem{Yao} Y. Yao, X. Xiao, L. Ge, C. P. Sun, Quantum coherence in multipartite systems, Phys. Rev. A 92, 022112 (2015).
\bibitem{Zurek} H. Ollivier, W. H. Zurek,  Quantum discord: A measure of the quantumness of correlations, Phys. Rev. Lett. 88, 017901 (2001).
\bibitem{Tan} K. C. Tan, H. Kwon, C.-Y.Park, H. Jeong, Unified view of quantum correlations and quantum coherence, Phys. Rev. A 94, 02 2329 (2016).
\bibitem{Kraft} T. Kraft, M. Piani,  Genuine correlated coherence, J. Phys. A: Math. Theor. 51, 414013 (2018).
\bibitem{Bhaskara}V. S Bhaskara, P. K Panigrahi,  Generalized concurrence measure for faithful quantification of multiparticle pure state entanglement using Lagrange's identity and wedge product, Quant. Inf. Process. 16, 118 (2017).
\bibitem{Tessier} T. E. Tessier, Complementarity relations for multi-qubit systems, Found. Phys. Lett. 18, 107 (2005).
\bibitem{Mark} J. A. Bergou, M. Hillery, \textit{Introduction to the Theory of Quantum Information Processing}, Springer, New York, 2013.
\bibitem{Messiah} A. Messiah, \textit{Quantum Mechanics}, Dover, New York, 2014.
\bibitem{Fano} U. Fano, Description of states in Quantum Mechanics by density matrix and operator techniques, Rev. Mod. Phys. 29, 74 (1957).
\bibitem{Jonas} J. Maziero, Hilbert-Schmidt quantum coherence in multi-qudit systems, Quantum Inf. Process. 16, 274 (2017).
\bibitem{Vedral} V. Vedral, M. B. Plenio, M. A., Rippin, P. L. Knight, Quantifying entanglement, Phys. Rev. Lett. 78, 2275 (1997).
\bibitem{Dhar} U. Singh, M. N. Bera, H. S. Dhar, A. K. Pati, Maximally coherent mixed states: Complementarity between maximal coherence and mixedness. Phys. Rev. A 91, 052115 (2015).
\bibitem{Vidal} W. D\"ur, G. Vidal, J. I. Cirac, Three qubits can be entangled in two inequivalent ways, Phys. Rev. A 62, 062314 (2000).
\bibitem{Coffman} V. Coffman,  J. Kundu, W. K. Wootters, Distributed entanglement, Phys. Rev. A 61, 052306 (2000).
\bibitem{Uhlmann} C. Eltschka, A. Osterloh, Siewert, A. Uhlmann, Three-tangle for mixtures of generalized GHZ and generalized W states, New J. Phys. 10, 043014 (2008).
\bibitem{Acin} A. Ac\'in, A. Andrianov, E. Jan\'e, R. Tarrach,  Three-qubit pure-state canonical forms, J. Phys. A: Math. Gen. 34, 6725 (2001).
\bibitem{Wit} A. Whitaker, \textit{Einstein, Bohr and the Quantum Dilemma}, Cambridge Univ. Press, 2nd Edition, 2006.
\end{thebibliography}
\end{document}